\documentclass[conference,letterpaper]{IEEEtran}
\IEEEoverridecommandlockouts

\usepackage[utf8]{inputenc} 
\usepackage[T1]{fontenc}
\usepackage{xurl}
\usepackage{ifthen}
\usepackage[cmex10]{amsmath} 

\usepackage{mleftright}       
\mleftright                   

\usepackage{graphicx}         
\usepackage{booktabs}         

\usepackage{comment}
\usepackage{balance}
\usepackage[url,hyperrefblack,notheorems,IEEEtran]{resea rch17} 
\usepackage{amsmath,amssymb,amsfonts,amsbsy}
\usepackage{tikz}
\usepackage{tikz-cd}

\usetikzlibrary{matrix,decorations.pathreplacing}
\pgfkeys{tikz/mymatrixenv/.style={decoration=brace,every left delimiter/.style={xshift=4pt},every right delimiter/.style={xshift=-4pt}}}
\pgfkeys{tikz/mymatrix/.style={matrix of math nodes,left delimiter=[,right delimiter={]},inner sep=1pt,row sep=0em,column sep=0em,nodes={inner sep=6pt}}}
\usepackage{amsthm} 
\newtheorem{theorem}{\mytheoremname}
\newtheorem{lemma}{\mylemmaname}
\newtheorem{corollary}{\mycorollaryname}

\newtheorem{proposition}{\mypropositionname}

\newtheorem{definition}{\mydefinitionname}
\newtheorem{remark}{\myremarkname}
\newtheorem{example}{\myexamplename}
\usepackage[capitalize]{cleveref}
\allowdisplaybreaks

\usepackage{array}
\usepackage{multirow}
\usepackage{enumitem} 
\usepackage{nicematrix} 
\usepackage[rightcaption]{sidecap} 
\newcommand*{\Scale}[2][4]{\scalebox{#1}{\ensuremath{#2}}} 

\newcommand{\Tan}{\textnormal{Tan}}
\newcommand{\Sch}{\textnormal{Sch}}
\newcommand{\Cay}{\textnormal{Cay}}
\newcommand{\Gr}{\mathcal{G}} 
\newcommand{\Gra}{\Gr_\textnormal{A}}
\newcommand{\Grb}{\Gr_\textnormal{B}}
\newcommand{\G}{\mathscr{G}} 
\newcommand{\Ga}{\mathscr{G}_\textnormal{A}}
\newcommand{\Gb}{\mathscr{G}_\textnormal{B}}
\newcommand{\Gs}{\mathscr{G}^\square}

\newcommand{\vv}{\mathcal{V}}
\newcommand{\ee}{\mathcal{E}}

\newcommand{\cev}[1]{\reflectbox{\ensuremath{\vec{\reflectbox{\ensuremath{#1}}}}}}
\newcommand{\aut}{\sigma}  
\newcommand{\cond}{the swapping condition} 
\usepackage{soul} 
\usepackage[colorinlistoftodos, textsize=footnotesize]{todonotes} 

\newboolean{EXTENDED_VERSION}
\setboolean{EXTENDED_VERSION}{true} 

\begin{document}
\title{Generalizing Quantum Tanner Codes} 



\author{%
    \IEEEauthorblockN{Olai \AA.~Mostad, Eirik Rosnes, and Hsuan-Yin Lin}
    \IEEEauthorblockA{Simula UiB, N--5006 Bergen, Norway}
    \IEEEauthorblockA{Emails: \{olai, eirikrosnes, lin\}{@}simula.no  }
}
\maketitle


\maketitle


\begin{abstract}
In this work, we present a generalization of the recently proposed quantum Tanner codes by Leverrier and Z{\'e}mor, which contains a construction of asymptotically good quantum low-density parity-check codes. Quantum Tanner codes have so far been constructed equivalently from groups, Cayley graphs, or square complexes constructed from groups. We show how to enlarge this to group actions on finite sets, Schreier graphs, and a family of square complexes, which is the largest possible in a certain sense. Furthermore, we discuss how the proposed generalization opens up the possibility of finding other families of asymptotically good quantum codes.
%
%
\end{abstract}


\section{Introduction}
\label{sec:introduction}

Quantum computers, which are based on the peculiarities of quantum mechanics, have been predicted to revolutionize several computing tasks for a long time, e.g., solving challenging problems that arise in chemistry and finance.   A quantum computer works by taking advantage of the quantum behavior of particles, which makes it possible to have superpositions of states. However, quantum computers are prone to errors due to the fragile nature of quantum states, in particular when the number of states grows. The use of quantum error-correction codes can mitigate the effect of such errors and hence make it possible to build large-scale quantum computers.
  
The existence of quantum error-correcting codes was first established independently by Shor and Steane in the mid-nineties~\cite{Shor95, Steane96_2}. CSS codes allowing to build a quantum code from two classical codes with the requirement that the dual of one should be contained in the other~\cite{CalderbankShor96_1, Steane96_1} were introduced shortly after and then followed by quantum stabilizer codes~\cite{Gottesman96, Calderbank98} which are in many ways analogous to classical linear codes. 
Since then, protecting quantum information has received considerable interest, and it was a long-standing open problem if asymptotically good 
quantum {low-density parity-check (LDPC)}
 error-correcting codes, i.e., 
{quantum LDPC} 
codes with a minimum distance growing linearly with the block length,
could exist. This was largely due to the CSS restriction that made it difficult to directly extend classical asymptotically good code constructions, and  was settled in a 2022 paper by Panteleev and Kalachev~\cite{PanteleevKalachev22_2}.
Subsequently, the construction in~\cite{PanteleevKalachev22_2} was modified and improved in~\cite{LeverrierZemor22_1}, resulting in a construction with an improved estimate of the minimum distance growth rate. Independently, very similar constructions have also answered a long-standing open question about locally testable classical codes~\cite{PanteleevKalachev22_2, DinurEvraLivneLubotzkyMozes22_1}. The recent renewed interest in quantum error correction has come due to recent progress in building intermediate-scale quantum computers with $300$-$1000$ qubits, enough to make them close to performing some tasks faster than state-of-the-art classical computers~\cite{Arute2019_1}.

\begin{figure}
    \centering\begin{tikzpicture}[scale=0.43]
        \begin{scope}[shift={(3cm,-5cm)}]
            \draw[fill=red, draw = black, fill opacity=0.4] (0,0) ellipse[x radius = 7, y radius = 2.6];
            \draw[fill=green, draw = black, fill opacity=0.4] (0,-0.6) ellipse [x radius = 6.2, y radius = 2];
            \draw[fill=blue, draw = black, fill opacity=0.4] (0,-1.2) ellipse[x radius = 5, y radius = 1.4];
            \draw[fill=orange, draw = black, fill opacity=0.4] (0,-1.8) ellipse[x radius = 4, y radius = 0.8];
            \node at (0,1.9) (A) {{General qTc}};
            \node at (0,0.7) (B) {{Our construction}};
            \node at (0,-0.7) (C) {{with symmetric $\set A, \set B$}};
            \node at (0,-1.9) (D) {qTc from \cite{LeverrierZemor22_1}};
        \end{scope}
    \end{tikzpicture}\caption{
    Our construction (green) generalizes the quantum Tanner codes  (qTc) from \cite{LeverrierZemor22_1} (brown). We also consider a slightly less general construction (blue), and in \cref{sec:equic-char} we show that there are codes not from our construction that also reasonably may be called general quantum Tanner codes (red).}
    \label{fig:venn-diagram}
    \vspace{-2ex}
\end{figure}

In this work, we propose a construction of quantum Tanner codes that can loosely be described as follows. Take two regular\footnote{A graph is called \emph{regular} if all its vertices have the same degree.} graphs on the same vertex set that commute, which can be thought of as their union having many four cycles. From the graphs, create a two-dimensional space of squares by filling in certain of these cycles. This type of space is fittingly called a square complex. The two diagonals in a square give rise to two new graphs, and by putting bits on the edges and parity-check constraints at the vertices of these graphs in a clever way, we obtain quantum codes. When applied to bipartite double covers of Cayley graphs, our construction gives the quantum Tanner codes from~\cite{LeverrierZemor22_1} (see~\cref{prop:prop_generalizing_qTanner}), illustrated by the brown region of \cref{fig:venn-diagram}. 

Our main technical result (\cref{thm:main}) gives a necessary and sufficient condition such that if one starts with a graph where this clever assignment is possible, then the graph can always be viewed as the graph of diagonals of a square complex, and this complex can always be made from commuting graphs. This condition puts us in the green region of \cref{fig:venn-diagram}. Such an assignment may also be possible without this condition, in which case the resulting codes still could reasonably be called general quantum Tanner codes. \cref{lemma:origclaim1} gives a condition for this, putting us in the red region of \cref{fig:venn-diagram}. 
%
There is potential to find other families of asymptotically good codes that do not come from Cayley graphs. The methods  in~\cite{LeverrierZemor23_1} can likely be adapted if one, e.g., has a non-Cayley Ramanujan Schreier graph commuting with a Cayley graph of two Ramanujan components, as discussed in~\cref{sec:commuting_Schreier_graphs}.
The codes we look at there are part of the blue region of \cref{fig:venn-diagram}, a subset of our codes that are easier to work with (see~\cref{sec:symm-labels}).

 \ifthenelse{\boolean{EXTENDED_VERSION}}{}{
{All proofs are omitted due to lack of space. 
They can be found in the extended version~\cite{MostadRosnesLin24_1sub}.
}}

\subsection{Notation}
\label{sec:notation}

Vectors  are denoted by bold letters, matrices by sans serif uppercase letters, 
%
and sets (and groups)  by calligraphic uppercase letters, e.g., $\bm a$, $\mat{A}$, and $\set{A}$, respectively. The neutral element of a group will be denoted by $1$, while $e$ is reserved for an edge in a graph. Linear codes, graphs,  
and square complexes are denoted by script uppercase letters, e.g., $\mathscr{C}$.  
%
A graph with vertex set $\set{V}$ and edge set $\set{E}$ is denoted by $\code{G}=(\set{V},\set{E})$, and may have parallel edges and self-loops unless stated otherwise. The edges incident to a vertex $v$ is called the local view of $v$ and denoted $\ee(v)$. We work with undirected graphs with an ordering on the edges and view the graphs as directed graphs (digraphs) with twice as many edges as we see fit. 
The disjoint union of sets $\set A, \set B$ is denoted by $\set A\sqcup \set B \eqdef \{(a,0), (b,1) : a\in \set A, b\in \set B\}$.
A linear code $\mathscr{C}$ of length $n$, dimension $k$, and minimum distance $d$ is sometimes referred to by 
$[n, k, d]$, and
its dual code is denoted $\mathscr{C}^\perp$. The binary field is denoted by $\Field_2$, the identity matrix of size $a$ by $\mat I_a$, the all-zero matrix (of arbitrary size) by $\mat 0$, and 
the transpose of a matrix by $\trans{(\cdot)}$.   
Standard order notation $\Theta(\cdot)$ is used for asymptotic results.
{Given a natural number $n$, we let $[n]\eqdef\{1,2,\ldots,n\}$.}

 

\section{Preliminaries}
\label{sec:preliminaries}

We recall some background on particular types of graphs, their (spectral) expansion, definitions of classical and quantum error-correcting codes, and the notion of a square complex.

\subsection{Graphs}
\label{sec:graphs}

\begin{definition}
  A labeling $\eta$ on a digraph $(\set{V}, \set{E})$ by elements of $\set{A}$ is a function $\eta: \set{E}\to \set{A}$. A digraph with a labeling is called a labeled digraph, and we say it is well-labeled if for every vertex $v\in \set{V}$ and label $a\in \set{A}$ there is exactly one edge starting at $v$ labeled by $a$ and exactly one edge ending in $v$ labeled by $a$. 
\end{definition}

A labeling on the local views of an undirected graph is equivalent to a labeling on the corresponding digraph. An edge $\begin{tikzcd}[cramped, sep=small]
    v \ar[r, dash, "e"] & {w}
\end{tikzcd}$
corresponds to a pair of directed edges $\begin{tikzcd}[cramped, sep=small]
    v \ar[r, "\vec {e}"] & {w,}
\end{tikzcd}$
$\begin{tikzcd}[cramped, sep=small]
    v  & {w,} \ar[l, "\cev{e}"{swap}]
\end{tikzcd}$
and we use the convention that $e$ has the label of $\vec{e}$ in the local view of $v$ and the label of $\cev{e}$ in the local view of $w$. We write $s(\vec e) = v = t(\cev e)$ and $t(\vec e) = w = s(\cev e)$,  where ``$s$'' and ``$t$'' indicate the source and target vertices of a directed edge, respectively. For bipartite graphs with vertex set $\vv = \vv_0\sqcup\vv_1$, we let $\vec{e}$ go from $\vv_0$ to $\vv_1$.
Given a (undirected) graph $(\vv, \ee)$, we write $\ee^{\textnormal{dir}}$ for the edges of the corresponding digraph.

Given a group $\set{G}$, we will call a subset $\set{A}\subseteq \set{G}$ \textit{symmetric} if $a^{-1}\in \set{A}$ for all $a\in \set A$.

\begin{definition}\label{def:cayley}
  Given a group $\set{G}$ and a symmetric subset $\set{A}\subseteq \set{G}$, the left Cayley graph $\textnormal{Cay}_\textnormal{l}(\set{G},\set{A})$ is the regular graph with vertex set $\set{G}$ and an edge $(g, g')$ if $g'=ag$ for an $a\in \set{A}$, in which case we label the edge by $a$ and $a^{-1}$ in the local view of $g$ and $g'$, respectively.\footnotemark[2]
\end{definition}
\footnotetext[2]{Cayley and Schreier graphs are often defined to have symmetric labeling sets and such that they have no self-loops or parallel edges.}

Right Cayley graphs $\textnormal{Cay}_\textnormal{r}(\set{G},\set{A})$ are defined similarly.

\begin{definition}
    A group action of $\Gr$ on $\vv$ is a function
    $
    {\varphi : \Gr\times \vv \to \vv}
    $
    such that $\varphi(1, v) = v$ and $\varphi(g , \varphi (h, v)) = \varphi((gh), v)$ for all $v\in \vv$ and $g,h\in\Gr$, where $1$ is the neutral element of $\Gr$.
\end{definition}

\ifthenelse{\boolean{EXTENDED_VERSION}}{
We write $gv$ for $\varphi(g,v)$ to simplify the notation.
A group action is called \textit{faithful} when $g=1$ is the only $g\in \Gr$ that acts trivially on $\vv$, i.e., such that $gv=v$ for all $v\in \vv$, \textit{free} when $g=1$ is the only $g\in\Gr$ such that $gv= v$ for some $v\in \vv$, \textit{transitive} when for any $v,w\in \vv$, there is a $g\in \Gr$ such that $g v = w$, and \textit{regular} when it is free and transitive.
}

\begin{definition}
    Given a group $\Gr$ acting on a set $\set{V}$ and a subset $\set{A}\subseteq \Gr$, the Schreier digraph $\textnormal{Sch}(\Gr, \set{V}, \set{A}) = (\vv, \ee_\textnormal A)$ 
    is the digraph with vertices $\set{V}$ and an edge $(v, v')\in \ee_\textnormal A$ labeled $a$ whenever there is an $a\in \set{A}$ mapping $v$ to $v'$ by the group action. A Schreier graph is constructed from a Schreier digraph by choosing a set of pairs $\begin{tikzcd}[cramped, sep=small]
    {v} \ar[r, shift left, ""] & {w} \ar[l, shift left, ""]
\end{tikzcd}$ of edges such that every (directed) edge is part of exactly one pair, and then making an edge $(v,w)$ for each pair on the above form.\footnotemark[2]
\end{definition}

For symmetric $\set A$, we will pair edges with inverse labels, as we do for Cayley graphs. Note that a directed edge can be paired with itself if it is a self-loop.
It is known that all regular graphs can be given the structure of a Schreier graph where $\set A$ is not necessarily symmetric.




\begin{remark}\label{rem:Sch_Cay}
    Schreier graphs are regular graphs with labeled local views so that the corresponding digraph is well-labeled. Cayley graphs are the Schreier graphs where the vertex set is the group $\Gr$, i.e., $\textnormal{Cay}_{\textnormal{l}}(\Gr,\set{A})=\Sch(\Gr,\Gr,\set{A})$  when $\Gr$ acts on the left. Both are labeled by the group elements $\set A \subseteq \Gr$.
\end{remark}

By Cayley's theorem~\cite{Cayley1878_1}, the elements of any group can be viewed as permutations of a set, turning the multiplication of elements in the group into a composition of functions. 
Going the other way, a set of permutations on a set $\set{V}$ will generate a group and define a directed Schreier graph of that group with vertex set $\set{V}$.
Concretely, we get a directed edge $v\to w$ labeled $\pi$ if $\pi(v)=w$.

With a stricter definition of Schreier graphs, most regular graphs are still Schreier.

\begin{proposition}[{\cite{GROSS77_1}}]
  \label{prop:most_graphs_are_Sch}
  All regular graphs of even degree can be given the structure of a Schreier graph with a symmetric labeling set. The same is true for graphs of odd degrees precisely when they have a perfect matching.\footnote[3]{A perfect matching is a set of edges where no edges share endpoints and all vertices are endpoints of edges in the set. We allow for self-loops in this set of edges.}
\end{proposition}

We will look at group actions of products of groups, i.e., commuting group actions (see \cref{rem:comm-actions}), and the following well-known facts will be useful.  

\begin{lemma}
  \label{lemma:commuting_perm_is_graph_hom}
   Two permutations $\pi_1,\pi_2:\vv\to\vv$ commute if and only if $\pi_2$ is a digraph homomorphism on the digraph $\G_1 = (\vv, \ee_1)$ where $\ee_1=\{(v,\pi_1(v)) : v\in \vv\}$, i.e., $(\pi_2(v), \pi_2(w))\in \ee_1$ whenever $(v,w)\in\ee_1$.   
  %
%
\end{lemma}

\ifthenelse{\boolean{EXTENDED_VERSION}}{
\begin{IEEEproof}
We have $(v,w)\in\ee_1$ if and only if $w=\pi_1(v)$,
and the permutations $\pi_1$ and $\pi_2$ commute if and only if $\pi_2(\pi_1(v)) = \pi_1(\pi_2(v))$ for every $v\in\vv$, so $(\pi_2(v), \pi_2(w))\in \ee_1$ for every edge $(v,w)\in\ee_1$ precisely when the permutations commute.
\end{IEEEproof}
}

\ifthenelse{\boolean{EXTENDED_VERSION}}{
\begin{proposition}
  \label{prop:trans_implies_regular}
  If a group $\Gr$ acts transitively and faithfully on a set $\set{X}$, then it acts regularly if and only if $\textnormal{Aut}_{\Gr}(\set{X})
  $ 
  acts transitively (and hence regularly),
  where $\textnormal{Aut}_{\Gr}(\set{X})$ is the group of functions $\aut:\set X \to \set X$ satisfying $\aut(gx)=g\aut(x)$ for all $x\in \set X, g\in \Gr$.
\end{proposition}
We add a proof for completeness. 
\begin{IEEEproof}
    Since $\Gr$ acts transitively, given $x,y\in \set X$, we can find a $g\in \Gr$ such that $gx=y$. Therefore, $\textnormal{Aut}_{\Gr}(\set{X})$ acts freely on $\set X$ because given $\aut\in\textnormal{Aut}_{\Gr}(\set{X})$ such that $\aut(x) = x$ for one $x\in \set X$, we have 
    \begin{IEEEeqnarray*}{c}
      \aut(y)=\aut(gx)=g\aut(x) {=gx = y}.
    \end{IEEEeqnarray*}
    So $\aut$ acts trivially on all ${y}\in \set X$, meaning $\aut = 1 \in \textnormal{Aut}_\Gr(\set X)$.
    
    Moreover, if $\Gr$ also acts freely on $\set X$, for given $x,y\in \set X$ we can define the function $\aut_{x,y}:\set X \to \set X$ by
    \begin{IEEEeqnarray*}{c}
      \aut_{x,y}(z) = g_{x,z}y
    \end{IEEEeqnarray*}
    for $z\in \set X$,
    where $g_{x,z}\in \Gr$ is the unique element mapping $x$ to $z$, i.e., $g_{x,z}x = z$. Now, observe that for any $h\in \Gr$ we have
    \begin{IEEEeqnarray*}{c}
      h\aut_{x,y}(z) = hg_{x,z}y \stackrel{(a)}{=} g_{x, hz}y = \aut_{x,y}(hz),
    \end{IEEEeqnarray*}
    where $(a)$ holds because $hg_{x,z}x = hz$ and $g_{x,hz}$ is the unique element such that $g_{x,hz}x=hz$. This indicates that $\aut_{x,y}$ is an element of $\textnormal{Aut}_{\Gr}(\set{X})$ by definition.
    We conclude that $\textnormal{Aut}_{\Gr}(\set{X})$ acts transitively on $\set X$ since
    \begin{IEEEeqnarray*}{c}
      \aut_{x,y}(x)=g_{x,x}y=1y = y,
    \end{IEEEeqnarray*}    
    where $1\in \Gr$ is the neutral element of the group.

    Conversely, we want to show that if $\textnormal{Aut}_{\Gr}(\set{X})$ acts transitively on $\set X$, then given $g \in \set G$ and some $x\in \set X$ such that $gx=x$, it implies $g=1$. Consider a $y\in \set X$ and let $\aut_{x, y}\in\textnormal{Aut}_{\Gr}(\set{X})$ such that $\aut_{x,y}(x)=y$. We have 
    \begin{IEEEeqnarray*}{c}
     g y = g \aut_{x,y}(x) \stackrel{(b)}{=} \aut_{x,y}(g x) = \aut_{x,y}(x) = y, 
    \end{IEEEeqnarray*}
    where $(b)$ is due to the definition of $\textnormal{Aut}_{\Gr}(\set{X})$. Hence, $g$ acts trivially on $\set X$. We assume that the group action is faithful, so it follows that $g=1$, and the action is also free.
\end{IEEEproof}}

Given two Schreier graphs $\G_{\textnormal{A}} = (\vv,\ee_\textnormal{A})$ and $\G_{\textnormal{B}} = (\vv,\ee_\textnormal{B})$  on the same vertex set, we will say they \textit{commute} if their defining permutations commute pairwise. 
That is, if $\Ga$ and $\Gb$ are labeled by $\eta_\textnormal{A}$ and $\eta_{\textnormal{B}}$, respectively, and we are given edges
$\begin{tikzcd}[cramped, sep=small]
    v_0 & v_1 \ar[l, "\cev e_1"{swap}] & v_2 \ar[l, "\cev e_2"{swap}] \ar[r, "\vec e_3"] & v_3 \ar[r, "\vec e_4"] & v_4
\end{tikzcd}$
such that $e_1, e_3\in \ee_\textnormal{A}$, $e_2,e_4\in \ee_{\textnormal{B}}$, $\eta_\textnormal{A}(\cev e_1) = \eta_\textnormal{A}(\vec e_3)$, and $\eta_\textnormal{B}(\cev e_2) = \eta_\textnormal{B}(\vec e_4)$, then $v_0=v_4$.
We will say they have \textit{overlapping edges} if there is a pair of vertices $v, w$ such that both graphs have at least one edge between $v$ and $w$.

\subsection{Graph Expansion}

By picking an order on the vertices of a graph $\G=(\vv,\ee)$, we get an adjacency matrix $\mat{M}^{\G}$ where $\mat{M}^{\G}_{ij}$ is the number of edges from the $j$-th vertex to the $i$-th vertex. 

Since $\mat{M}^{\G}$ is symmetric, it will have real eigenvalues $\lambda_1 \geq \cdots \geq \lambda_{|\vv|}$, where $\lambda_1 = \Delta$ when $\G$ is $\Delta$-regular, and $\lambda_{|\vv|} = -\Delta$ if and only if  it also is bipartite~\cite{Lubotzky1988_1}. For $\G$ connected and $|\vv|>2$, define 
$\lambda(\G) \triangleq \max\{\ecard{\lambda_i}\colon \lambda_i \neq \pm \Delta\}$. When $\G$ has several components (i.e. is disconnected), we set $\lambda(\G)=\Delta$.
This is a measure of the (spectral) expansion of the graph, and the graph is called \textit{Ramanujan} when $\lambda(\G) \leq 2\sqrt{\Delta - 1}$~\cite{Lubotzky1988_1}.





\subsection{Tanner Codes and Quantum CSS Codes}

Tanner codes were introduced by Tanner in~\cite{Tanner81_1} and famously give asymptotically good families of classical codes. Loosely speaking, the construction takes a graph, puts bits on the edges of the graph, and assigns a code to each vertex. We will be using a regular graph with the same code on every vertex. A choice of bits is then in the Tanner code if, for any vertex, the bits on the edges connected to the vertex are in the code assigned to it. Formally, we use the following definition, where the restriction of a vector $\vect c \in \Field_2^{|\ee|}$ defined on the edges $\ee$ of a graph to the local view of a vertex $v$ is denoted $\vect c_v$. Note that we assume an ordering on $\ee$ so that we may use $\Field_2^{|\ee|}$ instead of $\{ \ee \to \Field_2\}$ as our vector space.

\begin{definition}
  Let $\mathscr{C}$ be a linear code of length $\Delta$ and $\G=(\vv,\ee)$ be a $\Delta$-regular graph, possibly with parallel edges but without self-loops. We define the Tanner code on $\G$ and $\mathscr{C}$ as
  $\textnormal{Tan}(\G,\mathscr{C}) \triangleq \{\vect{c}\in\Field_2^{|\ee|}\colon\vect{c}_v\in\mathscr{C}\text{ for all } v\in \vv\}$.
\end{definition}

The definition assumes a well-labeling on $\G$. One may think of this as an order on each local view, where each order is independent of the other orderings.

{For} our main construction, the local code $\mathscr{C}$ will be the dual of a tensor product code.

\begin{definition}\label{def:tensor_code}
    Given linear codes $\mathscr{C}_\textnormal{A}, \mathscr{C}_\textnormal{B}$ of length $n_\textnormal{A}$ and $n_\textnormal{B}$, respectively, their tensor code $\mathscr{C}_\textnormal{A} \otimes \mathscr{C}_\textnormal{B}$ is defined as the set of $n_\textnormal{A} \times n_\textnormal{B}$ matrices with columns in $\mathscr{C}_\textnormal{A}$ and rows in $\mathscr{C}_\textnormal{B}$. 
\end{definition}

If $\mathscr{C}_\textnormal{A}$ and $\mathscr{C}_\textnormal{B}$ have parameters $[n_\textnormal{A}, k_\textnormal{A},  d_\textnormal{A}]$ and $[n_\textnormal{B}, k_\textnormal{B}, d_\textnormal{B}]$, respectively, then $\mathscr{C}_\textnormal{A} \otimes \mathscr{C}_\textnormal{B}$ has parameters $[n_\textnormal{A} n_\textnormal{B}, k_\textnormal{A} k_\textnormal{B}, d_\textnormal{A} d_\textnormal{B}]$. The dual code  $(\mathscr{C}_\textnormal{A} \otimes \mathscr{C}_\textnormal{B})^\perp$ is equal to $\mathscr{C}_\textnormal{A}^\perp \otimes \Field_2^{n_\textnormal{B}} + \Field_2^{n_\textnormal{A}} \otimes \mathscr{C}_\textnormal{B}^\perp$ and has minimum distance $\min(d_\textnormal{A}, d_\textnormal{B})$.

\begin{definition} \label{def:CSS}
    We say the classical codes $\mathscr{C}_0$ and $\mathscr{C}_1$ form a CSS code when $\mathscr{C}_0^\perp \subseteq \mathscr{C}_1$.
\end{definition}

If the classical codes $\mathscr{C}_0$ and $\mathscr{C}_1$ have parity-check matrices $\mat{H}_0$ and $\mat{H}_1$, respectively, \cref{def:CSS} is equivalent to $\mat{H}_0 \trans{\mat{H}_1} = \mat{0}$. CSS codes were introduced in~\cite{CalderbankShor96_1}, where they show that the classical codes $\mathscr{C}_0$ and $\mathscr{C}_1$ can be used to construct good quantum error-correcting codes.

The dimension $k$ of a CSS code where the classical codes are of length $n$ is
  $k = \dim(\mathscr{C}_0 \setminus \mathscr{C}_1^\perp) = \dim \mathscr{C}_0 + \dim \mathscr{C}_1 - n$,
and the minimum distance $d$ of the quantum CSS code can be given as the minimum of
  $d_\textnormal{X}=\min_{\vect{c}\in\mathscr{C}_0 \setminus \mathscr{C}_1^\perp}|\vect{c}|$ and 
  $d_\textnormal{Z} = \min_{\vect{c}\in\mathscr{C}_1\setminus\mathscr{C}_0^\perp}|\vect{c}|$.

A CSS code $(\mathscr{C}_0, \mathscr{C}_1)$ is called a \textit{quantum LDPC} code when both codes $\mathscr{C}_0$ and $\mathscr{C}_1$ are defined by sparse parity-check matrices. For families of codes, we require that the columns and rows of the parity-check matrices have weight at most $\Delta$, for some constant $\Delta$ independent of the code length $n$.
  A code family is called \textit{asymptotically good} if it has parameters $[n, k = \Theta(n), d = \Theta(n)]$.

\subsection{Square Complexes}

We will need the notion of square complexes, normally defined as two-dimensional cube complexes, a particular type of CW complex~\cite{Whitehead1949_1}. We will use the following definition, which is equivalent for our purposes.

\begin{definition}
    A square complex $\set X = (\vv, \ee, \set Q)$ is a triple of sets such that $(\vv,\ee)$ is a graph and the elements of $\set Q$ are of the form $((v_1,v_2),(v_1,v_3),(v_2,v_4),(v_3,v_4))\in \ee^{\times 4}$.


\end{definition}






\section{Proposed Generalized Construction}
\label{sec:main-results}

We start by giving an example of commuting non-Cayley Schreier graphs. Then, we
give a construction of quantum LDPC codes that generalize the quantum Tanner codes of~\cite{LeverrierZemor22_1} and can take these Schreier graphs as input. We compare the two constructions and characterize the new one in three different ways.

\subsection{Example}
\label{sec:Petersen-graph}

The Petersen graph~\cite{HoltonSheehan93_1}, pictured to the left in Fig.~\ref{fig:Petersen graph}, is known to be a non-Cayley graph, and can be considered as two $5$-cycles joined in a certain way. 
Let
\begin{IEEEeqnarray*}{c}
  \mat{C}_5 = \left[ \begin{smallmatrix}
    0 & 1 & 0 & 0 & 1\\
    1 & 0 & 1 & 0 & 0\\
    0 & 1 & 0 & 1 & 0\\
    0 & 0 & 1 & 0 & 1\\
    1 & 0 & 0 & 1 & 0
    \end{smallmatrix} \right] \quad\textnormal{and}\quad
    \mat{C}'_5 = \left[\begin{smallmatrix}
    0 & 0 & 1 & 1 & 0\\
    0 & 0 & 0 & 1 & 1\\
    1 & 0 & 0 & 0 & 1\\
    1 & 1 & 0 & 0 & 0\\
    0 & 1 & 1 & 0 & 0
    \end{smallmatrix} \right]
\end{IEEEeqnarray*}
denote different adjacency matrices for a $5$-cycle (vertices labeled by $1,2,\ldots,5$ for $\mat{C}_5$ and vertices labeled by $1',\ldots,5'$ for $\mat{C}_5'$ in the left graph in Fig.~\ref{fig:Petersen graph}). In a certain basis, the Petersen graph has the adjacency matrix
\begin{IEEEeqnarray*}{c}
  \mat{M}_\textnormal{A}=
  \left[
    \begin{array}{c|c}
      \mat{C}_5 & \mat{I}_5 \\
      \hline
      \mat{I}_5 & \mat{C}'_5
    \end{array}
  \right], 
  \text{ commuting with }
  \mat{M}_{\textnormal B}=
  \left[
    \begin{array}{c|c}
    \mat{C}_5 & \mat{0} \\
      \hline
      \mat{0} & \mat{C}_5
    \end{array}
  \right]\rlap{,}
\end{IEEEeqnarray*}
which is the adjacency matrix of the second graph depicted in Fig.~\ref{fig:Petersen graph}, so the pair of graphs give an example of a non-Cayley graph commuting with a $2$-component graph when labeled as in the figure.
%
By reordering the vertices, one can also write the adjacency matrices as 
\begin{IEEEeqnarray*}{c}
  \mat{M}_{\textnormal A}=
  \left[
    \begin{array}{c|c}
      \mat{C}_5 & \mat{P} \\
      \hline
      \trans{\mat{P}} & \mat{C}_5
    \end{array}
  \right] \quad \text{and} \quad 
  \mat{M}_{\textnormal B}=
  \left[
    \begin{array}{c|c}
      \mat{C}_5 & \mat{0} \\
      \hline
    \mat{0} & \mat{C}'_5
    \end{array}
  \right],
\end{IEEEeqnarray*}
for a certain permutation matrix $\mat{P}$.

The two graphs have overlapping edges and different degrees. This is unwanted for our applications and may be remedied, for example, in the following way. First, add self-loops to all vertices of the second graph to make their degrees equal. Then, take two copies of the resulting graph, and use the bipartite double cover of the Petersen graph (the Desargues graph~\cite{HoltonSheehan93_1}), as explained in \cref{sec:old-construction}. If one in the end also wants both graphs to be bipartite on the same partition of vertices, one may take the bipartite double cover of both resulting graphs.

\begin{figure}[t!]
  \centering
\begin{tikzpicture}[scale=1,auto, swap,
circ/.style={draw,shape=circle,fill=black, inner sep=0.3ex},
lab/.style={shape=circle, font=\scriptsize}]
\path (0,1) node[circ] (p0) {}
(-0.95105651629,0.30901699437) node[circ] (p1) {}
(-0.58778525229,-0.80901699437) node[circ] (p2) {}
(0.58778525229,-0.80901699437) node[circ] (p3) {}
(0.95105651629,0.30901699437) node[circ] (p4) {}
(0,2) node[circ] (q0) {}
(-0.95105651629*2,0.30901699437*2) node[circ] (q1) {}
(-0.58778525229*2,-0.80901699437*2) node[circ] (q2) {}
(0.58778525229*2,-0.80901699437*2) node[circ] (q3) {}
(0.95105651629*2,0.30901699437*2) node[circ] (q4) {};
\node[above=of p0, xshift=-8pt, yshift=-8pt] {\scriptsize{$1$}};
\node[left=of p1, xshift=1.5mm, yshift=4.5mm] {\scriptsize{$2$}};
\node[below=of p2, xshift=-22pt, yshift=12pt] {\scriptsize{$3$}};
\node[below=of p3, xshift=22pt, yshift=12pt] {\scriptsize{$4$}};
\node[above=of p4, xshift=1.06cm, yshift=-8mm] {\scriptsize{$5$}};
\node[below=of q0, yshift=4mm, xshift=-2mm] {\scriptsize{$1'$}};
\node[right=of q1, xshift=-5.7mm, yshift=-4.5mm] {\scriptsize{$2'$}};
\node[above=of q2, xshift=3mm, yshift=-5mm] {\scriptsize{$3'$}};
\node[above=of q3, xshift=-3mm, yshift=-5mm] {\scriptsize{$4'$}};
\node[below=of q4, xshift=-7.2mm, yshift=9mm] {\scriptsize{$5'$}};
\draw[-Stealth] (p0) -- node[near start, yshift=-6pt] {\scriptsize{$a_1$}} (p2);
\draw[-Stealth] (p1) -- node[near start, xshift=4pt, yshift=2pt] {\scriptsize $a_1$} (p3);
\draw[-Stealth] (p2) -- node[near start, xshift=-3pt] {\scriptsize $a_1$} (p4);
\draw[-Stealth] (p3) -- node[near start] {\scriptsize $a_1$} (p0);
\draw[-Stealth] (p4) -- node[near start] {\scriptsize $a_1$} (p1);
\draw[-Stealth] (q0) -- node[lab] {\scriptsize $a_1$} (q1);
\draw[-Stealth] (q1) -- node[lab] {\scriptsize $a_1$} (q2);
\draw[-Stealth] (q2) -- node[lab, yshift=4pt] {\scriptsize $a_1$} (q3);
\draw[-Stealth] (q3) -- node[lab] {\scriptsize $a_1$} (q4);
\draw[-Stealth] (q4) -- node[lab] {\scriptsize $a_1$} (q0);
\draw[] (p0) -- node[near start] {\scriptsize $a_0$} (q0);
\draw[] (p1) -- node[yshift=-2pt, xshift=-4pt] {\scriptsize $a_0$} (q1);
\draw[] (p2) -- node[swap, yshift=4pt] {\scriptsize $a_0$} (q2);
\draw[] (p3) -- node[yshift=4pt] {\scriptsize $a_0$} (q3);
\draw[] (p4) -- node[swap, yshift=-2pt, xshift=4pt] {\scriptsize $a_0$} (q4);
\end{tikzpicture}
\begin{tikzpicture}[scale=1.0,auto, swap, node distance=10m,
circ/.style={draw,shape=circle,fill=black, inner sep=0.3ex},
lab/.style={shape=circle, font=\scriptsize}]
\path (0,1) node[circ] (p0) {}
(-0.95105651629,0.30901699437) node[circ] (p1) {}
(-0.58778525229,-0.80901699437) node[circ] (p2) {}
(0.58778525229,-0.80901699437) node[circ] (p3) {}
(0.95105651629,0.30901699437) node[circ] (p4) {}
(0,2) node[circ] (q0) {}
(-0.95105651629*2,0.30901699437*2) node[circ] (q1) {}
(-0.58778525229*2,-0.80901699437*2) node[circ] (q2) {}
(0.58778525229*2,-0.80901699437*2) node[circ] (q3) {}
(0.95105651629*2,0.30901699437*2) node[circ] (q4) {};
\draw[-Stealth] (p0) -- node[] {\scriptsize $b_0$} (p1);
\draw[-Stealth] (p1) -- node[lab] {\scriptsize $b_0$} (p2);
\draw[-Stealth] (p2) -- node[lab, yshift=3pt] {\scriptsize $b_0$} (p3);
\draw[-Stealth] (p3) -- node[lab] {\scriptsize $b_0$} (p4);
\draw[-Stealth] (p4) -- node[lab] {\scriptsize $b_0$} (p0);
\draw[-Stealth] (q0) -- node[lab] {\scriptsize $b_0$} (q1);
\draw[-Stealth] (q1) -- node[lab] {\scriptsize $b_0$} (q2);
\draw[-Stealth] (q2) -- node[lab, yshift=4pt] {\scriptsize $b_0$} (q3);
\draw[-Stealth] (q3) -- node[lab] {\scriptsize $b_0$} (q4);
\draw[-Stealth] (q4) -- node[lab] {\scriptsize $b_0$} (q0);
\end{tikzpicture}
\vspace{-22pt}
\caption{The Petersen graph is shown to the left. On the right is a Schreier graph commuting with the Petersen graph. They are labeled by $\set A = \{a_0,a_1,a_2\}$ and $\set B = \{b_0,b_1\}$,  respectively, where $a_0^{-1}=a_0$, $a_1^{-1}=a_2$,  and $b_0^{-1}=b_1$.}
\vspace{-11.1pt}
\label{fig:Petersen graph}
\end{figure}

\subsection{New Construction}
\label{sec:new-construction}

\subsubsection{General Case}

Let $\G_{\textnormal A} = \Sch(\Gra, \vv, \set A) = (\vv, \ee_\textnormal A)$ and $\G_{\textnormal B} = \Sch(\Grb, \vv, \set B) = (\vv, \ee_\textnormal B)$ be (non-directed) commuting $\Delta$-regular Schreier graphs with no overlapping edges and a chosen partition $\vv = \vv_0 \sqcup \vv_1$ for which both graphs are bipartite.
We treat the graphs as digraphs and call the labelings they have in virtue of being Schreier graphs $\eta_\textnormal A : \ee_{\textnormal A}^{\textnormal{dir}} \to \set A$ and $\eta_\textnormal B : \ee_\textnormal B^{\textnormal{dir}} \to \set B$, respectively. 
Furthermore, assume that if two vertices $v,w$ are connected by an edge in $\Ga$, then the {sets of pairs of ``inverses'' for the two vertices are equal, meaning $\{ (\eta_{\textnormal B}(\vec e), \eta_{\textnormal B}(\cev e)) : e \in \ee_{\textnormal{B}}(v)\} = \{ (\eta_{\textnormal B}(\vec e), \eta_{\textnormal B}(\cev e)) : e \in \ee_{\textnormal{B}}(w)\}$, and vice versa when swapping the role of $\textnormal A$ and $\textnormal B$.}


From the commuting graphs $\Ga, \Gb$, we may construct a square complex $\set X$ with vertices $\vv$, edges $\set E_\textnormal{A}\cup \set E_\textnormal{B}$, and for each $v\in \vv_0$, $a\in \set A$, and $b\in \set B$, a square $(e_1,e_2,e_3,e_4) \in \ee_\textnormal A \times \ee_\textnormal A \times \ee_\textnormal B \times \ee_\textnormal B$ given by \eqref{eq:square} below. We illustrate it by the square on the left when $t(\vec e_1) = (w, 1)$, $t(\vec e_3)= (w',1)$, $s(\vec e_2) = (v',0)$, $\eta_\textnormal{A}(\vec e_1) = a$, and $\eta_\textnormal{B}(\vec e_3) = b$.
\newline
\begin{minipage}[c]{0.32\linewidth}
  \begin{tikzcd}[baseline=7ex]
    (w',1) \ar[r, "a", "\cev e_2"{swap}] & (v',0)
    \\
    (v,0) \ar[r, "a", "\vec e_1"{swap}] \ar[u, "b", "\vec e_3"{swap}] \ar[ur, phantom, ""{description}] & (w,1) \ar[u, "b", "\cev e_4"{swap}]
  \end{tikzcd}
  \end{minipage}\hfill
\begin{minipage}[c]{0.65\linewidth}
\begin{IEEEeqnarray}{c}
\begin{cases}
    s(\vec e_1)=s(\vec e_3)=(v,0),
    \\
    \eta_\textnormal A (\vec e_1)=\eta_\textnormal A (\cev e_2),\, s(\cev e_2)=t(\vec e_3),
    \\
    \eta_\textnormal B (\vec e_3)=\eta_\textnormal B (\cev e_4),\, s(\cev e_4)=t(\vec e_1).
\end{cases}
\label{eq:square}
\end{IEEEeqnarray}
\end{minipage}
The squares $(e_1, e_2, e_3, e_4)$ and $(e_2, e_1, e_4, e_3)$ are identified.
We refer to $\set X$ as the Schreier complex on $\Ga$ and $\Gb$, and denote its set of squares by $\set Q$.

We define $\Gs_0 = (\vv_0, \set{E}^\set{Q}_0)$ as the $\Delta^2$-regular graph with vertices $\vv_0$ and an edge $(v, v')\in\set{E}^\set{Q}_0$ labeled by $(\eta_\textnormal A (\vec e_1), \eta_\textnormal B (\vec e_3)) =(a,b)\in \set A\times \set B$ in the local view of $v$ and $(\eta_{\textnormal{A}}(\vec e_2), \eta_{\textnormal{B}}(\vec e_4))$ in the local view of $v'$, for each square on the form~\eqref{eq:square}. 
Similarly, we let $\Gs_1 = (\vv_1, \set E^\set Q_1)$ be the $\Delta^2$-regular graph with vertices $\vv_1$ and an edge $(w, w')$ labeled $(\eta_\textnormal A (\cev e_1), \eta_\textnormal B (\cev e_4))$ in the local view of $w$ and $(\eta_\textnormal A (\cev e_2), \eta_\textnormal B (\cev e_3))$ in the local view of $w'$ for each square on the form~\eqref{eq:square}.

\ifthenelse{\boolean{EXTENDED_VERSION}}{
\begin{remark}
    {A bipartite graph $\G=(\vv_0\sqcup \vv_1, \ee)$ gives rise to graphs $(\vv_0, \ee^2_0)$ and $(\vv_1, \ee^2_1)$, called the bipartite halves of $\G$, where we have an edge $(v,v')$ in $\ee^2_i$ for each (unordered) pair of edges $(v, w),(w,v')$ in $\G$ with $w\in\vv_{i+1}$ 
    (addition mod 2).}
    The graphs $\Gs_0$ and $\Gs_1$ are the subgraphs of the two bipartite halves of $\Ga \cup \Gb$ where all edges come from pairs $e_0,e_1$ with $e_0\in\Ga, e_1\in\Gb$.
\end{remark}
}

\begin{definition}\label{def:main}
    Given graphs as above and classical codes $\mathscr{C}_\textnormal{A}$, $\mathscr{C}_\textnormal{B}$ of length $\Delta$, define $\mathscr{C}_0$ and $\mathscr{C}_1$ as the Tanner codes
    \begin{IEEEeqnarray*}{c}
      \Scale[1.0]{\mathscr{C}_0 = \Tan(\Gs_0, (\mathscr{C}_\textnormal{A}\otimes\mathscr{C}_\textnormal{B})^\perp),\,
      \mathscr{C}_1 = \Tan(\Gs_1, (\mathscr{C}_\textnormal{A}^\perp\otimes\mathscr{C}_\textnormal{B}^\perp)^\perp)}.
    \end{IEEEeqnarray*}
\end{definition}

See \hyperref[app]{Appendix} for a concrete construction of parity-check matrices for the codes $\mathscr{C}_0$ and $\mathscr{C}_1$ from \cref{def:main}.

\cref{prop:CSS-code} below is proved similarly to the corresponding statement in~\cite{LeverrierZemor22_1}. 

\begin{proposition}
  \label{prop:CSS-code}
  The  codes $\mathscr{C}_0$ and $\mathscr{C}_1$ form a CSS code which is also a quantum LDPC code.
\end{proposition}

\ifthenelse{\boolean{EXTENDED_VERSION}}{
Before proving the proposition, we recall some vocabulary used for a Tanner code $\mathscr{D} = \Tan((\vv, \ee), \mathscr{C}^\perp)$. Let a $\mathscr{C}$-generator for $\mathscr{D}$ be a vector of $\Field_2^{|\ee|}$ which is equal to a codeword of $\mathscr{C}$ on the local view of some vertex $v$, and zero elsewhere. With this terminology, $\mathscr{D}$ is the space orthogonal to all $\mathscr{C}$-generators.

\begin{IEEEproof}
Without loss of generality, let $\set A = \{a_1,\dots, a_\Delta\}$ and $\set B = \{b_1,\dots, b_\Delta\}$. 
    Next, let $\vect v$ be a $(\mathscr{C}_\textnormal{A}\otimes \mathscr{C}_\textnormal{B})$-generator for $\mathscr{C}_0$, and let $\vect w$ be a $(\mathscr{C}_\textnormal{A}^\perp\otimes \mathscr{C}_\textnormal{B}^\perp)$-generator for $\mathscr{C}_1$, supported on $\ee^\set Q_0(v)$ and $\ee^\set Q_1(w)$, respectively. If these supports overlap, then $(v, 0)$ and $(w, 1)$ are connected by either edges in $\Ga$ or edges in $\Gb$. Without loss of generality, suppose the edges are from $\Ga$ and are labeled by $\{a_i: i\in \set I\}$ in $\ee_\textnormal A (v,0)$ and by $\{a_{i'}: i'\in \set I'\}$ in $\ee_\textnormal A (w,1)$, for some subsets $\set I, \set I' \subseteq [\Delta]$ of the same size.
    Then the overlap between $\ee^\set Q_0(v)$ and $\ee_1^\set Q(w)$ will be labeled by $\{ (a_i,b_j) : i\in \set I, j\in [\Delta]\}$ in $\ee^\set Q_0(v)$ and by $\{ (a_{i'},b_j) : i'\in \set I', j\in [\Delta]\}$ in $\ee^\set Q_1(w)$.
    By the definition of the tensor code (see \cref{def:tensor_code}), the generator $\vect v$ is now a codeword of $\mathscr{C}_\textnormal{B}$ on the set $\{ (a_i,b_j) :j\in [\Delta]\}$ for each $i \in \set{I}$, while $\vect w$ is a codeword of $\mathscr{C}_\textnormal{B}^\perp$ on the set $\{ (a_{i'},b_j) : j \in [\Delta]\}$ for each  $i'\in \set I'$, so $\vect v$ and $\vect w$ are orthogonal.
    Recall that the definition of $\vect v$ and $\vect w$ implies that $\code C_0^\perp$ is the span of all such $\vect{v}$ and $\mathscr{C}_1$ is the space orthogonal to all such $\vect{w}$. We conclude that $\code C_0^\perp \subseteq \mathscr{C}_1$, and hence the codes form a CSS code.

    For the LDPC part, we can keep the degree $\Delta$ of the graphs involved fixed while letting the number of vertices they have grow.
\end{IEEEproof}
}


\begin{remark}
\label{rem:comm-actions}
    Two Schreier graphs commute precisely when the group actions $\set \Gra \times \vv\to \vv$ and $\set \Grb  \times \vv\to \vv$ defining them form a group action $\set \Gra \times \set \Grb \times \vv \to \vv$. Hence, without loss of generality, we {could} define our construction using a group action $\set \Gra\times \set \Grb  \times \vv\to \vv$ and subsets $\set A\subseteq \Gr_\textnormal{A}, \set B\subseteq \Gr_\textnormal{B}$ instead of the commuting Schreier graphs $\Sch(\Gr_\textnormal{A}, \vv, \set A)$ and $\Sch(\Gr_\textnormal{B}, \vv, \set B)$.
\end{remark}

\subsubsection{Symmetric Labeling Set}
\label{sec:symm-labels}

The construction used in \cref{def:main} can be somewhat simplified when the labeling sets are symmetric. In this case, the inverse of each label is well-defined. When the graphs involved are not already bipartite (with respect to the same partition of vertices), we can make them so by using the bipartite double cover of the graphs, simplifying it further. In this case, $\Gs_0 = \Gs_1$.

\subsection{Connection With Previous Constructions}\label{sec:old-construction}

To create commuting graphs $\Ga, \Gb$, one may start with a group $\Gr$ and two symmetric subsets $\set A, \set B \subseteq \Gr$. Then the Cayley graphs $\Ga = \Cay_\textnormal{l}(\Gr, \set A)$ and $\Gb = \Cay_\textnormal{r}(\Gr, \set B)$ will commute because group multiplication is associative. Our construction on the bipartite double covers of these graphs is equivalent to the approach used to create quantum Tanner codes so far~\cite{LeverrierZemor22_1}.


Our assumption that the graphs $\Ga, \Gb$ have no overlapping edges plays the same role as the total no-conjugacy (TNC)  condition for the quantum Tanner codes defined on groups, which states that $ag\neq gb$ for all $g\in \Gr, a\in \set A, b\in \set B$.
It ensures that $v$ and $v'$ in~\eqref{eq:square} are different so that there are no self-loops in $\Gs_0, \Gs_1$.
Many authors {use}
``the quadripartite construction'' to avoid dealing with the TNC condition.

In our setup, the quadripartite construction corresponds to the regular construction on two copies of one of the graphs and the bipartite double cover of the other.
In other words, for graphs with adjacency matrices $\mat{M}_\textnormal{A}$ and $\mat{M}_\textnormal{B}$, use the graphs with adjacency matrices 
\begin{IEEEeqnarray*}{c}
\left[
\begin{array}{c|c}
    \mat 0 & \mat{M}_\textnormal{A}\\
     \hline
    \mat{M}_\textnormal{A} & \mat 0
\end{array}
\right]\quad\textnormal{and}\quad
\left[
\begin{array}{c|c}
    \mat{M}_\textnormal{B} & \mat 0\\
     \hline
    \mat 0 & \mat{M}_\textnormal{B}
\end{array}
\right].
\end{IEEEeqnarray*}
It can easily be seen that the two graphs still commute after this step when the obvious labeling is chosen.
In the case of Cayley graphs, one may equivalently swap the group $\Gr$ for $\Gr \times \mathbb F_2$, and use $\set A' = \{(a,1): a \in \set A\}$ and $\set B' = \{(b,0): b \in \set B\}$.
This means that using the quadripartite construction is quite restrictive when looking for concrete finite-length examples. 

The following proposition should be clear when comparing our construction with the one from~\cite{LeverrierZemor22_1}.

\begin{proposition}[]
  \label{prop:prop_generalizing_qTanner}
  Let $\Gr$ be a group with generating symmetric subsets $\set A$ and $\set B$ of size $\Delta$ satisfying the TNC condition and not containing the neutral element, and let $\mathscr{C}_\textnormal{A}, \mathscr{C}_\textnormal{B}$ be codes of length $\Delta$.
  Then, our construction applied to the bipartite double covers of the graphs $\textnormal{Cay}_\textnormal{l}(\Gr, \set A), \textnormal{Cay}_\textnormal{r}(\Gr, \set B)$ and the codes $\mathscr{C}_\textnormal{A}, \mathscr{C}_\textnormal{B}$ gives the same CSS code as the construction from~\cite{LeverrierZemor22_1} applied to $\Gr, \set A, \set B, \code C_\textnormal{A}, \code C_\textnormal{B}$.
\end{proposition}

From \cref{prop:most_graphs_are_Sch}, most regular graphs can be used to construct quantum Tanner codes. However, to have freedom when choosing the other graph, the automorphism group of the graph should be large. Moving away from Cayley graphs means getting a smaller automorphism group, see \cref{sec:commuting_Schreier_graphs}.


\subsection{Equivalent Characterizations}
\label{sec:equic-char}

It is natural to ask when a square complex can give CSS codes the way left-right Cayley complexes and our square complexes described in~\cref{sec:old-construction} do, namely, by changing which diagonal of the squares that determines their endpoints when viewed as edges.
We now turn to prove that these are precisely the square complexes that can be made from two commuting Schreier graphs (see \cref{cor:square_complexes_are_Sch}). Along the way, we present another view of quantum Tanner codes (\cref{lemma:origclaim1}), and show how our construction fits in (\cref{thm:main}).


{In \cref{lemma:origclaim1}, we consider $\Delta^2$-regular Schreier graphs. The local views are taken to be labeled by $\set A \times \set B$ for sets $\set A, \set B$ of size $\Delta$
    so that the local view $\set{E}(v)$ of a vertex $v$ can be viewed as a matrix $\mat{E}(v)$ of size $\Delta\times\Delta$, where its entry $\mat{E} (v)_{a,b}=e\in\set{E}$ when {$e$ is labeled $(a,b)$ in $\ee(v)$}. 
    Thus, the Tanner codes and the rows and columns of local views are well-defined.}

\begin{lemma} 
    \label{lemma:origclaim1}
    Let $\mathscr{G}_0 = (\vv_0,\ee_0)$ and $\mathscr{G}_1 = (\vv_1,\ee_1)$ be $\Delta^2$-regular Schreier graphs labeled by $\set A \times \set B$ for sets $\set A, \set B$ of size $\Delta$ such that $|\vv_0|=|\vv_1|$, and let $\psi:\ee_0\to \ee_1$ be the bijection given by the order on the edges. Then, (i) and (ii) are equivalent.
    \begin{itemize} 
      \item[(i)] The Tanner codes $\mathscr{C}_0 = \Tan(\G_0, (\mathscr{C}_\textnormal{A}\otimes\mathscr{C}_\textnormal{B})^\perp)$ and $\mathscr{C}_1 = \Tan(\G_1, (\mathscr{C}_\textnormal{A}^\perp\otimes\mathscr{C}_\textnormal{B}^\perp)^\perp)$ form a CSS code for all classical codes $\mathscr{C}_\textnormal{A}, \mathscr{C}_\textnormal{B}$ of length $\Delta$.

      \item[(ii)] 
      For any vertices $v\in \vv_0,w\in\vv_1$, either $\set U \eqdef \psi(\ee_0(v))\cap\ee_1(w)= \emptyset$, or $\set U$ forms one or more rows or columns in the local views matrix $\mat{E}_1(w)$, such that each row (column)  is mapped by $\psi^{-1}$ to a row (column) in $\mat{E}_0(v)$.
    \end{itemize}
  \end{lemma}
  We find it reasonable to call any codes $\code C_0, \code C_1$ constructed as in $(i)$ \textit{general quantum Tanner codes}, so these codes fit in the red region of \cref{fig:venn-diagram}. 
  Note that $\Delta^m$-regular graphs with $m$ local codes $\code C_{\textnormal{A}_1},\ldots,\code C_{\textnormal{A}_m}$ also are of interest and could share this name. However, we restrict ourselves to the case $m=2$. 
\ifthenelse{\boolean{EXTENDED_VERSION}}{	
  \begin{IEEEproof}
Without loss of generality, let $\set A = \{a_1,\dots, a_\Delta\}$ and $\set B = \{b_1,\dots, b_\Delta\}$. 
    Next, let $v,w$ be vertices such that $\psi(\ee_0(v))\cap\ee_1(w) = \set U \neq \emptyset$.
    Recall that $\code C_0^\perp$ is the span of all $(\code C_\textnormal{A}\otimes \code C_\textnormal{B})$-generators for $\code C_0$, while $\code C_1$ is the space orthogonal to the $(\code C_\textnormal{A}^\perp\otimes \code C_\textnormal{B}^\perp)$-generators for $\code C_1$.
    Since $(i)$ holds, i.e., $\code{C}_0^\perp\subseteq\code{C}_1$, this tells us that the $(\code C_\textnormal{A}\otimes \code C_\textnormal{B})$-generators for $\code C_0$ are orthogonal to the $(\code C_\textnormal{A}^\perp\otimes \code C_\textnormal{B}^\perp)$-generators for $\code C_1$, 
    meaning a codeword of $(\mathscr{C}_\textnormal{A}\otimes\mathscr{C}_\textnormal{B})$ restricted to $\psi^{-1}(\set U) \subseteq \ee_0(v)$ must be orthogonal to a codeword of $(\mathscr{C}_\textnormal{A}^\perp\otimes\mathscr{C}_\textnormal{B}^\perp)$ restricted to $\set U$ regardless of $\code {C}_\textnormal{A}$ and $\code{C}_\textnormal{B}$.
    With our choice of local codes, this implies that $\set U$ is labeled by a set of the form $\{(a_i,b_j) : i\in [\Delta], j\in \set J\}$ (or $\{(a_i,b_j) : i\in \set I, j\in [\Delta]\}$) in $\ee_1(w)$, {for some $\set J$ (or $\set I)\subseteq [\Delta]$}, and $\psi^{-1}(\set U)$ is labeled by $\{(a_i,b_{j'}) : i\in [\Delta], j' \in \set J'\}$ (or $\{(a_{i'},b_j) : i' \in \set I', j\in [\Delta]\}$) in $\ee_0(v)$, {for some $\set J'$ (or $\set I')\subseteq [\Delta]$ of the same size as $\set J$ (or $\set I$)}.
    In the first case, the edges must ``line up'' so that the edge labeled $(a_i, b_{j'})$ in $\psi^{-1}(\set U)$ is mapped to the edge labeled $(a_i, b_j)$ in $\ee_1(w)$ by $\psi$ (mapping columns). In the other case $\psi$ must map the edge labeled $(a_{i'}, b_j)$ to the edge labeled $(a_i, b_j)$ (mapping rows).

  
    To prove that $(ii)$ implies $(i)$ we proceed as follows. Let $\vect{v}$ be a $(\mathscr{C}_\textnormal{A}\otimes\mathscr{C}_\textnormal{B})$-generator for $\mathscr{C}_0$, and let $\vect{w}$ be a $(\mathscr{C}_\textnormal{A}^\perp\otimes\mathscr{C}_\textnormal{B}^\perp)$-generator for $\mathscr{C}_1$, supported at $\ee_0(v)$ and $\ee_1(w)$, respectively. We want to show that $\vect{v}$ and $\vect{w}$ are orthogonal to each other so that $\code C_0$ and $\code C_1$ form a CSS code. Suppose, without loss of generality, that $\set U$ is labeled by $\{ (a_i,b_j) : i\in \set I, j\in [\Delta]\}$ in $\ee_1(w)$, {for some $\set I \subseteq [\Delta]$}, and $\psi^{-1}(\set U)$ is labeled by $\{ (a_{i'},b_j) : i'\in \set I', j\in [\Delta]\}$ in $\ee_0(v)$, for some $\set I'\subseteq[\Delta]$ of the same size as $\set I$, mapped by $\psi$ as above.
    The generator $\vect v$ is now a codeword of $\mathscr{C}_\textnormal{B}$ on the set of edges labeled by $\{ (a_{i'},b_j) : j\in [\Delta]\}$ in $\ee_0(v)$ for each $i'\in \set{I}'$, while $\vect w$ is a codeword of $\mathscr{C}_\textnormal{B}^\perp$ on the image of each of these sets of edges by $\psi$,
    and it follows that  $\vect v$ and $\vect w$ are orthogonal.
\end{IEEEproof}
}

\cref{thm:main} gives a condition that restricts the codes constructed in \cref{lemma:origclaim1} from the red region of \cref{fig:venn-diagram} to the green region.
	
\begin{theorem}
  \label{thm:main}
    Let $\G_0$, $\G_1$, and $\psi$ be as in \cref{lemma:origclaim1}.
    Then, (i) and (ii) of \cref{lemma:origclaim1} are equivalent to (iii) below
    if and only if 
    for each of the two labels of an edge $\psi$ fixes, the labels share one index in the labeling set $\Delta\times \Delta$.
  \begin{itemize}
  \item[(iii)] There exist $\Ga=\Sch(\Gr_\textnormal A, \vv, \set A)$, $\Gb=\Sch(\Gr_\textnormal B, \vv, \set B)$ for $\vv\eqdef\vv_0\sqcup\vv_1$ such that  $\Gs_i$ constructed from $\Ga, \Gb$ as in \cref{sec:new-construction}, equals $\G_i$ for $i=0,1$. 
  \end{itemize}
\end{theorem}

We refer to the condition on $\psi$ mentioned in \cref{thm:main} as \textit{\cond}, where the name comes from \cref{lemma:swapping_cond} below.
We need this lemma and one more (\cref{lemma:labels_loc_inv}) to prove \cref{thm:main}, so we state them before turning to the proof of the theorem.

\begin{lemma}
  \label{lemma:swapping_cond}
  Let $\G_0$, $\G_1$, and $\psi$ as in \cref{lemma:origclaim1} satisfying \cond \ be given.
  Then, an edge in $\mathscr{G}_0$ corresponding to $(a,b)$ and $(a',b')$ in its local views must be mapped by $\psi$ to an edge corresponding to $(a,b')$ and $(a',b)$ in its local views.
\end{lemma}
 
    \begin{IEEEproof}
        Condition $(ii)$ of  \cref{lemma:origclaim1} implies that $a, b, a', b'$ all need to be possible to fix individually, so the only other option is $\psi$ fixing the labels, in which case \cond \ of \cref{thm:main} tells us that $a'=a$ and $b'=b$. 
    \end{IEEEproof}	

\begin{lemma}\label{lemma:labels_loc_inv}
    Let $\G_0$, $\G_1$, and $\psi$ as in \cref{lemma:origclaim1} satisfying \cond \ be given, and let $(v,v')\in \ee_0$ be an edge labeled $(a,b)$ in $\ee_0(v)$ and $(a',b')$ in $\ee_0(v')$. The edges in the local view of $v$ with $a$ as the first entry of their label will then have $a'$ as the first entry of the label in their other local view. Similarly, the edges of $\ee_0(v)$ having $b$ as the second entry of their label will have $b'$ as the second entry of their label in their other local view.
\end{lemma}

\begin{IEEEproof}
    The image of a set of edges $\{e_i\}$ forming a row in $\mat E_0(v)$ has to form a row (in the same order as before) in some local view after being mapped by $\psi$. For $\{\psi(e_{i})\}$ to form a row in this way, the second entry of their label must be the same as it is in $\mat E_0(v)$, and the first entry of the label must be common for all the edges. By \cond, this entry is precisely the first entry of the labels of the edges in $\{e_i\}$ in their other local view (i.e., not the local view of $v$, but of their other endpoint).
\end{IEEEproof}

\begin{IEEEproof}
  [Proof of~\cref{thm:main}]
  We have already shown that $(iii)$ implies $(i)$ in \cref{prop:CSS-code}, and $(i)$ and $(ii)$ are equivalent by \cref{lemma:origclaim1}. \cref{thm:main} says that $(ii)$ implies $(iii)$ precisely when \cond \ is satisfied. As a diagram, what we want to show is that the gray implication arrow below holds precisely when the swapping condition is satisfied.

$$
\begin{tikzcd}[column sep=40]
    {(i)} \ar[d, Leftrightarrow, "\textnormal{\cref{lemma:origclaim1}}"{swap, xshift=-2}] 
    & {(iii)} \ar[l, Rightarrow, "\textnormal{\cref{prop:CSS-code}}"{swap, yshift=2}]
    \\
    {(ii)} \ar[ur, color=gray, Rightarrow, "\textnormal{Only assuming the}"{swap, color=black, yshift=2}, "\textnormal{swapping condition}"{swap, color=black, yshift=-5, xshift=0}]
\end{tikzcd}
$$
    We will first show the only-if part of the theorem by showing that all $\psi$ coming from our construction will satisfy \cond. Then, we show the if-part of the theorem by constructing $\G_\textnormal{A}$ and $\G_{\textnormal{B}}$ as in $(iii)$ assuming \cond. Lemmas \ref{lemma:swapping_cond} and \ref{lemma:labels_loc_inv} will be used to show that the construction is well-defined.
  
  We now prove the only-if part of the theorem by showing that \cond \ will be satisfied whenever $(iii)$ is satisfied by the following argument.
An edge in $\Gs_0$ coming from the square 
$$
\begin{tikzcd}
    (w',1) \ar[r, dash, "e_2"] & (v',0)
    \\
    (v,0) \ar[r, dash, "e_1"{}] \ar[u, dash, "e_3"{}] & (w,1) \ar[u, dash, "e_4"{swap}]
  \end{tikzcd}
$$
corresponds to the edge in $\Gs_1$ coming from the same square. Recall that we in a bipartite graph with vertices $\vv_0 \sqcup \vv_1$ let $\eta(\vec e)$ denote the label of $e$ in the local view of its endpoint with second index $0$. In $\Gs_0$ it has the labels $(\eta(\vec e_1), \eta(\vec e_3))$ and $(\eta(\vec e_2), \eta(\vec e_4))$, and in $\Gs_1$ it has the labels $(\eta(\vec e_2), \eta(\vec e_3))$ and $(\eta(\vec e_1), \eta(\vec e_4))$. So if one of the two labels the edge has in $\Gs_0$ is equal to a label of the edge in $\Gs_1$, this implies that either $\eta(\vec e_1)=\eta(\vec e_2)$ or $\eta(\vec e_3)=\eta(\vec e_4)$, and if both labels are equal we have both $\eta(\vec e_1)=\eta(\vec e_2)$ and $\eta(\vec e_3)=\eta(\vec e_4)$. This means that the $\psi$ we get from our construction in \cref{sec:main-results} satisfies \cond, proving the only-if part of \cref{thm:main}.


Next, we prove the if-part of the theorem. In particular, 
we show that $(ii)$ implies $(iii)$ given \cond, i.e., that \cond \ is not only necessary to satisfy $(iii)$, but also sufficient.

\ifthenelse{\boolean{EXTENDED_VERSION}}{


 We do this by constructing a square complex in the following way.
 Given an edge $(v,v')$ labeled $(a,b)$ in $\ee_0(v)$ and $(a',b')$ in $\ee_0(v')$ mapped by $\psi$ to an edge $(w, w')$, \cref{lemma:swapping_cond} tells us that we may assume the edges are labeled $(a',b)$ in $\ee_1(w)$ and $(a,b')$ in $\ee_1(w')$. For each such edge, we make the square
 \vspace{-4pt}
 \[
   \begin{tikzcd}[ampersand replacement=\&]
     (w',1) \ar[r, "{a}", shift left] \ar[d, "{b'}", shift left] \& (v',0) \ar[d, "{b'}", shift left] \ar[l, "{a'}", shift left] 
     \\
     (v,0) \ar[r, "a", shift left] \ar[u, "b", shift left] \ar[ur, phantom, "q"{description}] 
     \& (w,1)\rlap{,} \ar[u, "{b}", shift left] \ar[l, "a'", shift left]
   \end{tikzcd}
 \vspace{-1pt}
 \]
 meaning four vertices, four edges labeled in each direction as indicated, and one square glued onto them.

Below is an example picture, where the fat blue edge $(v_1,v_4)$ is mapped to the fat red edge $(w_3,w_4)$, giving rise to the black square $(((v_1,0),(w_3,1)),\allowbreak ((v_4,0),(w_4,1)),\allowbreak ((v_1,0),(w_4,1)),\allowbreak ((v_4,0),(w_3,1)))$.
$$
\begin{tikzpicture}[scale=1,auto, swap,
            circ/.style={draw,shape=circle,fill=black, inner sep=0.3ex}]
        \path 
    (0,0) node[circ] (0) {}
    (2,0) node[circ] (1) {}
    (2,2) node[circ] (2) {}
    (0,2) node[circ] (3) {}
    (5.5,-0.41421356237) node[circ] (4) {}
    (7.41421356237-0.5,1) node[circ] (5) {}
    (5.5,2.41421356237) node[circ] (6) {}
    (4.5-0.41421356237,1) node[circ] (7) {};
    \draw[color=blue] (0) -- (1) -- (2) -- (3) -- (0);
    \draw[color=blue, very thick] (0) --  node[yshift=-10pt, swap] {$(a,b)$} node[yshift=10pt, swap] {$(a',b')$} (3);
    \draw[color=red] (4) -- (5) -- (6) -- (7) -- (4);
    \draw[color=red, very thick] (6) -- node[yshift=10pt, xshift=14] {$(a',b)$} node[yshift=-10pt, xshift=-5] {$(a,b')$} (7);
    \draw[->] (2.5,1) -- node[yshift=0pt, swap] {$\psi$} (3.25,1);
    \node[left=of 0, xshift=30] {$v_1$};
    \node[right=of 1, xshift=-30] {$v_2$};
    \node[right=of 2, xshift=-30] {$v_3$};
    \node[left=of 3, xshift=30] {$v_4$};
    \node[right=of 4, xshift=-30] {$w_1$};
    \node[right=of 5, xshift=-30] {$w_2$};
    \node[right=of 6, xshift=-30] {$w_3$};
    \node[left=of 7, xshift=30] {$w_4$};
\end{tikzpicture}
$$%
$$
\begin{tikzpicture}[scale=1,auto, swap,
            circ/.style={draw,shape=circle,fill=black, inner sep=0.3ex}]
        \path 
    (0,0) node[circ] (0) {}
    (2,0) node[circ] (1) {}
    (2,2) node[circ] (2) {}
    (0,2) node[circ] (3) {}
    (1,-0.41421356237) node[circ] (4) {}
    (2.41421356237,1) node[circ] (5) {}
    (1,2.41421356237) node[circ] (6) {}
    (-0.41421356237,1) node[circ] (7) {};
    \fill[color=black!20] (3.center) -- (7.center) -- (0.center) -- (6.center) -- (3.center);
        \path 
    (0,0) node[circ] (0) {}
    (2,0) node[circ] (1) {}
    (2,2) node[circ] (2) {}
    (0,2) node[circ] (3) {}
    (1,-0.41421356237) node[circ] (4) {}
    (2.41421356237,1) node[circ] (5) {}
    (1,2.41421356237) node[circ] (6) {}
    (-0.41421356237,1) node[circ] (7) {};
    \draw[color=blue] (0) -- (1) -- (2) -- (3) -- (0);
    \draw[color=red] (4) -- (5) -- (6) -- (7) -- (4);
    \draw[color=black] (3) --  node[yshift=2, xshift=5] {$a'$} node[yshift=-10pt, xshift=-1] {$a$} (7) -- node[yshift=10pt, xshift=-2] {$b'$} node[yshift=-3pt, xshift=2] {$b$} (0) -- node[yshift=10pt, xshift=1] {$a'$} node[yshift=-10pt, xshift=-5] {$a$} (6) -- node[yshift=-3] {$b'$} node[yshift=2pt, xshift=10] {$b$} (3);
    \node[left=of 0, xshift=32, yshift=-10] {$(v_1,0)$};
    \node[right=of 1, xshift=-30] {$(v_2,0)$};
    \node[right=of 2, xshift=-30] {$(v_3,0)$};
    \node[left=of 3, xshift=32, yshift=10] {$(v_4,0)$};
    \node[right=of 4, xshift=-30] {$(w_1,1)$};
    \node[right=of 5, xshift=-30] {$(w_2,1)$};
    \node[right=of 6, xshift=-30] {$(w_3,1)$};
    \node[left=of 7, xshift=30] {$(w_4,1)$};
\end{tikzpicture}
$$

 Next, equal vertices are identified, and so are edges with equal labels between equal vertices.
From \cref{lemma:labels_loc_inv}, it follows that 
given a vertex $v$ and an element $x\in \set A \cup \set B$, we will have exactly one edge in the local view of $v$ labeled $x$ after this identification. 
We obtain a square complex $\set X$ with vertices $\vv_0\sqcup\vv_1$, where the underlying graph is bipartite with local views labeled by $\set A \cup \set B$, and squares on the above form.
%
%
  This underlying graph may be split into 
  two Schreier graphs with labeling set $\set A$ and $\set B$, respectively.
  It is clear that we get back the same square complex $\set X$ when doing the construction described in \cref{sec:new-construction} on these two Schreier graphs, and that the diagonals of the squares in this square complex are precisely the edges of $\G_0$ and $\G_1$ that correspond to each other.
  Hence, $(iii)$ holds. 
  %
\end{IEEEproof}	

\begin{corollary}
\label{cor:square_complexes_are_Sch}
    All square complexes that give CSS codes using the construction from \cref{sec:new-construction}, can be constructed from a pair of commuting Schreier graphs as in \cref{sec:new-construction}.
\end{corollary}
\ifthenelse{\boolean{EXTENDED_VERSION}}{
\begin{IEEEproof}
    {Condition $(i)$ of  \cref{lemma:origclaim1} is satisfied since the square complex gives rise to a quantum Tanner code, and since $\psi$ is constructed by changing the diagonal of a square complex, \cond \ is satisfied by the proof of the only-if part of  \cref{thm:main}. Therefore, the result follows from \cref{thm:main}.}
\end{IEEEproof}
}

We give an easy example showing the existence of cases where condition $(i)$ of  \cref{lemma:origclaim1} holds but \cond \ of  \cref{thm:main} does not. This indicates that the red region of \cref{fig:venn-diagram} is strictly larger than the green one, even when restricting ourselves to the local codes we use here.
\begin{example}
  \label{ex:red_nonempty}
 \begin{figure}
     \centering
     %
     %
{\begin{tikzpicture}[scale=1.6, auto, swap,
circ/.style={draw,shape=circle,fill=black, inner sep=0.3ex},
lab/.style={shape=circle, font=\scriptsize}]
\path (0,0) node[circ] (0) {}
(2,0) node[circ] (1) {}
(1,1.73205080757) node[circ] (2) {};
\draw[-Stealth] (0) to [] node[yshift=0pt, swap] {\scriptsize{$(a_0,b_0)$}} (1);
\draw[-Stealth] (1) to [] node[swap, yshift=-5pt, xshift=8pt] {\scriptsize{$(a_0,b_0)$}} (2);
\draw[-Stealth] (2) to [] node[yshift=10pt, swap] {\scriptsize{$(a_0,b_0)$}} (0);
\draw[-Stealth] (0) to [bend right=40] node[] {\scriptsize{$(a_0,b_1)$}} (1);
\draw[-Stealth] (1) to [bend right=40] node[] {\scriptsize{$(a_0,b_1)$}} (2);
\draw[-Stealth] (2) to [bend right=40] node[] {\scriptsize{$(a_0,b_1)$}} (0);
\end{tikzpicture}}
     \caption{{The depicted graph is considered in \cref{ex:red_nonempty}. It is $4$-regular with edges depicted as arrows. An edge has the indicated label in the local view of the source of the arrow depicting it. An edge labeled $(a_0, b_0)$ in one of its local views is labeled $(a_1, b_0)$ in the other, and similarly the ``inverse'' of $(a_0, b_1)$ is $(a_1, b_1)$.}}
     \label{fig:example-red_nonempty}
 \end{figure}
    {\cref{fig:example-red_nonempty} shows a $4$-regular graph $\G$ with local views labeled by $\{a_0, a_1\}\times \{b_0, b_1\}$ {such that an edge labeled $(a_0,b_i)$ in one local view is labeled $(a_1,b_i)$ in the other, for $i\in \{0,1\}$.}
     Using $\G_0=\G_1=\G$ as the depicted graph and $\psi=\textnormal{id}$, where $\textnormal{id}$ denotes the identity mapping, $(ii)$ is satisfied by the following observation. Two overlapping local views can either be local views of the same vertex or of neighboring vertices. In the first case, it can be viewed as either two columns or two rows, and in the second, it is the first row of one of the local views and the second row of the other. However, $\psi$ fixes both labels of each edge, and only the second index is shared in the two labels of each edge, meaning that \cond \ of  \cref{thm:main}  is not satisfied.}
 \end{example}
}

\begin{remark}
\cref{ex:red_nonempty} can be viewed as an example of a construction where each edge in a $\Delta$-regular graph $\G$ is swapped for $\Delta$ parallel edges given a labeling such that $\psi=\textnormal{id}$ satisfies condition $(ii)$ of \cref{lemma:origclaim1}. 
Hence, these $\Delta^2$-regular graphs can be used to make general quantum Tanner codes.
The graph may also be constructed as $\Gs_0 = \Gs_1$ using $\Ga = \G$ and a graph of only (self-inverse) self-loops as $\Gb$.
However, not every labeling of the graph that would satisfy (ii) may be constructed using the construction from \cref{sec:new-construction}, as \cref{ex:red_nonempty} shows.
\end{remark}


\section{Asymptotically Good Quantum Codes}
\label{sec:commuting_Schreier_graphs}

In this section, we discuss how our proposed construction 
might be used to create new families of asymptotically good codes. The discussion assumes that the graphs are not already bipartite, and will be made so by taking their bipartite double cover. Then, since $\Gs_0$ and $\Gs_1$ are isomorphic, we will write $\Gs$ to lighten notation. We start by stating Proposition \ref{prop:graphs_are_disc} below, which gives an obstruction for when a pair of commuting graphs can be non-Cayley.

Note that \cref{prop:graphs_are_disc} considers Cayley graphs that allow for multiplicities in the set $\set A$ of \cref{def:cayley}, where Cayley graphs are defined. Also, recall that we do not demand $\set A$ to be generating the group, so we allow Cayley graphs to have more than one connected component.

\begin{proposition}
  \label{prop:graphs_are_disc}
  If $\Ga$ and $\Gb$ commute, then they are either both Cayley graphs or one of them has more than one component.
\end{proposition}
\ifthenelse{\boolean{EXTENDED_VERSION}}{
\begin{IEEEproof}
    Let $\Ga$ and $\Gb$ be connected commuting graphs.
  According to Lemma~\ref{lemma:commuting_perm_is_graph_hom}, when a permutation $\pi_\textnormal{b}$ commutes with a set of permutations $\set A$, this is equivalent to the permutation being a graph homomorphism of the graph $\Ga = \Sch(\Gr(\set A), \vv, \set A)$, where $\Gr(\set A)$ is the group generated by $\set A$. The graph $\Ga$ is connected, so the action of $\Gr(\set A)$ on $\vv$ is transitive.
  Moreover, since the graphs commute, all $\pi_\textnormal{b}$ must be elements of $\textnormal{Aut}_{\Gr(\set A)}(\vv)$, and because the graph $\Gb$ is connected, $\textnormal{Aut}_{\Gr(\set A)}(\vv)$ therefore acts transitively.
Assuming the action of $\Gr(\set A)$ on $\vv$ is also faithful, Proposition \ref{prop:trans_implies_regular} now tells us that $\Gr(\set A)$ must act regularly on $\vv$.
  
  If the group action of $\Gr(\set A)$ on $\vv$ is not faithful, then, because the labeling sets {may, without loss of generality, be assumed to} be symmetric for a graph commuting with a connected graph, the group elements that act the same way on $\vv$ can be viewed as several copies of the same group element, making the group action faithful.

  A Schreier graph $\Sch(\Gr, \vv, \set A)$ defined by a regular group action $\Gr\times \vv \to \vv$ must be a Cayley graph in the sense of \cref{rem:Sch_Cay} by the following argument. Fix a vertex $v_0\in \vv$, and let $g_v\in \Gr$ for $v\in \vv$ be the unique group element such that $g_v v_0 = v$. Now, give the set $\vv$ a group structure by letting $v\cdot w = g_v g_w$. The group $(\vv, \cdot)$ is now isomorphic to $\Gr$, and this isomorphism takes the group action to group multiplication.
\end{IEEEproof}
}


At first glance, it might seem like Proposition~\ref{prop:graphs_are_disc} tells us there is no hope of finding asymptotically good quantum codes using the methods from~\cite{LeverrierZemor23_1}. After all, $\lambda(\G)=\Delta$ for a $\Delta$-regular graph $\G$ with more than one component, which is as large as it can get. However, as already seen in~\cref{sec:old-construction}, we can get good codes even in this case, as one of the graphs will have two components when using the quadripartite construction. This stems from the fact that $\G^\square$ and the components of $\Ga$ and $\Gb$ may have a small $\lambda$.

Since the adjacency matrices $\mat{M}_\textnormal{A}$ and $\mat{M}_\textnormal{B}$ are symmetric and commute, they are simultaneously diagonalizable. Therefore,  $\mat{M}_\textnormal{A} + \mat{M}_\textnormal{B}$ and $\mat{M}_\textnormal{A} \mat{M}_\textnormal{B}$, which are the adjacency matrices of, respectively, $(\vv, \set E_\textnormal{A}\cup \set E_\textnormal{B})$ and $\G^\square$, have eigenvalues the sums and products, respectively, of the eigenvalues of $\mat{M}_\textnormal{A}$ and $\mat{M}_\textnormal{B}$.

We know that the all-ones vector $\vect u$ will correspond to $\lambda_1=\Delta$ for any regular graph, and for a graph with two components commuting with a connected graph, the other eigenvector corresponding to this eigenvalue that is also an eigenvector for the other graph will have to be $(\vect u,-\vect u)$.

Let $\mat{M}_\textnormal{A} = \left[
\begin{array}{c|c}
  \mat{A}_1 & \mat{A}_2\\
  \hline
  \trans{\mat{A}_2} & \mat{A}_3
\end{array}
\right]$ and $\mat{M}_\textnormal{B} = \left[
\begin{array}{c|c}
  \mat{B}_1 & 0\\
  \hline
  0 & \mat{B}_2
\end{array}
\right]$
be the adjacency matrices of $\Ga$ and $\Gb$, respectively. We demand that $\mat M_\textnormal{A} \mat M_\textnormal{B} = \mat M_\textnormal{B} \mat M_\textnormal{A}$, which means that the product is a symmetric matrix. These two products are
\begin{IEEEeqnarray*}{c}
\left[
\begin{array}{c|c}
    \mat{A}_1 \mat{B}_1 & \mat{A}_2 \mat{B}_2\\
     \hline
    \trans{\mat{A}_2} \mat{B}_1 & \mat{A}_3 \mat{B}_2
\end{array}
\right]
\quad\textnormal{and}\quad 
\left[
\begin{array}{c|c}
    \mat{B}_1 \mat{A}_1 & \mat{B}_1 \mat{A}_2\\
     \hline
    \mat{B}_2\trans{\mat{A}_2} & \mat{B}_2 \mat{A}_3
\end{array}
\right] \rlap{,}  
\end{IEEEeqnarray*}
which are equal if and only if we have the relations
$\mat{A}_2 \mat{B}_2 = \mat{B}_1 \mat{A}_2$, $\mat{A}_1 \mat{B}_1 = \mat{B}_1 \mat{A}_1$, and  $\mat{A}_3 \mat{B}_2 = \mat{B}_2 \mat{A}_3$.  Assuming that $\mat{B}_1$, $\mat{B}_2$, $\mat{A}_1$, and $\mat{A}_3$ correspond to connected graphs, \cref{prop:graphs_are_disc} now tells us they all have to be Cayley graphs.
\Cref{lemma:commuting_perm_is_graph_hom} tells us that $\mat{A}_2$ either is the $\mat{0}$ matrix or a sum of permutation matrices that represent  graph isomorphisms between the two components of $\Gb$, so the two components of $\Gb$ are equal up to a rearrangement of the vertices since we assume $\Ga$ is connected.
All this fits what we saw in the example of~\cref{sec:Petersen-graph}.

\begin{remark}\label{rem:comm_blocks}
     $\Ga$ and $\Gb$ commute, so if $a\in \set A$ takes $v$ to $w$, then all vertices in the component of $\Gb$ containing $v$ will be mapped by $a$ to vertices in the same component of $\Gb$ as $w$. This means that each $a\in \set A$ contributes to exactly one $\mat A_i$.
\end{remark}

If we 
let {$\alpha$} be the regularity of the graph corresponding to $\mat{A}_1$, then the eigenvalue of $\mat{M}_\textnormal{A}$ corresponding to $(\vect u,-\vect u)$ is {$2\alpha-\Delta$}. So, when the weights of the rows in $\mat{A}_1$ and $\mat{A}_2$ are equal, then  {$\lambda(\G^\square) \leq (2\sqrt{\Delta-1})^2 = 4(\Delta-1)$} when $\Ga$ and the two components of $\Gb$ are Ramananujan graphs, because $\lambda_2=\Delta$ for $\Gb$ is multiplied with $0$.

We end by pointing at a possible way to create a connected Schreier graph $\Ga$ that commutes with a Cayley graph $\Gb$ with two components. Let $\Ga$ be a Cayley graph on the above form, and let $\mat P$ be a permutation matrix of the same size as $\mat{A}_3$. If $\trans{\mat{P} \mat{A}_3 \mat{P}}$ commutes with $\mat{B}_2$, then the matrix
\begin{IEEEeqnarray}{c}
  \left[
    \begin{array}{c|c}
      \mat{A}_1 & \mat{A}_2\\
      \hline
      \trans{\mat{A}_2} & \mat{P} \mat{A}_3 \trans{\mat{P}}
    \end{array}
  \right]
  \label{eq:block_matrix}
\end{IEEEeqnarray}
will still commute with $\mat M_\textnormal{B}$.
We can ensure this by choosing $\mat{P}$ so that $\mat{P} \mat{A}_3 \trans{\mat{P}}$ is the adjacency matrix of a left Cayley graph of the same group that $\mat B_2$ is a right Cayley graph of. For example, given a Cayley graph $\Cay_{\textnormal{l}}(\Gr,\set{A})$ and an automorphism $\sigma$ on $\Gr$, then $\sigma$ also is an isomorphism between $\Cay_{\textnormal{l}}(\Gr,\set{A})$ and $\Cay_{\textnormal{l}}(\Gr,\set{A}')$, where $\set{A}'=\{\sigma (a) : a\in \set{A}\}$. A clever choice of $\sigma$ should make \eqref{eq:block_matrix} the adjacency matrix of a non-Cayley Schreier graph.\looseness-1

\balance
\appendix
\label{app}

We give a short description of how parity-check matrices for $\code C_0 = \Tan(\Gs_0, (\mathscr{C}_\textnormal{A}\otimes\mathscr{C}_\textnormal{B})^\perp)$ and $\code C_1 = \Tan(\Gs_1, (\mathscr{C}_\textnormal{A}^\perp\otimes\mathscr{C}_\textnormal{B}^\perp)^\perp)$ from \cref{def:main} may be constructed.

We assume an ordering on the edges of $\Gs_i = (\vv_i, \ee^\set Q_i)$ such that $e_0\in \ee^\set Q_0$ and $e_1\in \ee^\set Q_1$ appear at the same place of the two ordered sets when they come from the same square of the Schreier complex, making the sets of edges $\ee_i^{\set Q} = \{e_{i,1}, \dots, e_{i,n}\}$. Here, $n$ is the length of the code, which is equal to 
$|\set E_i^{\set Q}| = |\vv_0| \Delta^2 /2$. 
We also order the vertices $\vv_i=\{v_{i,1},\dots,v_{i,|\vv_0|}\}$ and write $\set A = \{a_1, \dots, a_\Delta\}$ and $\set B = \{b_1,\dots, b_\Delta\}$.

For an $m_{\textnormal{A}}\times n_{\textnormal{A}}$ matrix $\mat A$ and an $m_{\textnormal{B}}\times n_{\textnormal{B}}$ matrix $\mat B$, their \textit{Kronecker product} is the $m_{\textnormal{A}}m_{\textnormal{B}}\times n_{\textnormal{A}}n_{\textnormal{B}}$ matrix
$$
\mat A\otimes \mat B =
\begin{bmatrix}
    a_{11}\mat B & \dots & a_{1n_{\textnormal{A}}}\mat B\\
    \vdots & \ddots & \vdots \\
    a_{m_{\textnormal{A}}1}\mat B & \dots & a_{m_{\textnormal{A}}n_{\textnormal{A}}}\mat B
\end{bmatrix}\rlap{.}
$$
The codewords of the tensor code from \cref{def:tensor_code} are defined as $n_\textnormal{A}\times n_\textnormal{B}$ matrices $\mat C$. When we flatten them into vectors of length $n_\textnormal{A} n_\textnormal{B}$ by stacking the rows of the matrices, we get codewords of the form $\bm c = (c_{11}, \dots , c_{1n_{\textnormal{B}}}, \dots , c_{n_{\textnormal{A}}1} , \dots , c_{n_{\textnormal{A}}n_{\textnormal{B}}})$. It is clear that the Kronecker product $\mat H_\textnormal{A}\otimes \mat H_\textnormal{B}$ becomes a parity-check matrix for 
the flattened version of
$(\code C_\textnormal{A}^\perp\otimes \code C_{\textnormal{B}}^\perp)^\perp = \code C_\textnormal{A} \otimes \Field_2^{n_\textnormal{B}} + \Field_2^{n_\textnormal{A}} \otimes \code C_\textnormal{B}$
if $\mat H_\textnormal{A}$ and $\mat H_\textnormal{B}$ are parity-check matrices for $\code C_\textnormal{A}$ and $\code C_\textnormal{B}$, respectively.


To see this, recall that given vectors $\vect u =(u_1,\ldots,u_{n_\textnormal{A}}) \in \Field_2^{n_{\textnormal{A}}}$ and $\vect v = (v_1,\ldots,v_{n_\textnormal{B}}) \in \Field_2^{n_{\textnormal{B}}}$, we have $\vect u \otimes \vect v = (u_{1}  v_1, \dots, u_{1}  v_{n_\textnormal{B}}, \dots ,u_{n_\textnormal{A}} v_1, \dots,  u_{n_\textnormal{A}} v_{n_\textnormal{B}})$.
A flattened codeword $\vect c$ of $\code C_\textnormal{A} \otimes \Field_2^{n_\textnormal{B}} + \Field_2^{n_\textnormal{A}} \otimes \code C_\textnormal{B}$ can be written on the form $\vect c = \sum_{j=1}^{n_\textnormal{B}} \vect {{c_{\textnormal{A}}}}_j \otimes \vect e_j + \sum_{i=1}^{n_\textnormal{A}} \vect e_i \otimes \vect {{c_{\textnormal{B}}}}_i$, where $\vect {{c_{\textnormal{A}}}}_j\in \code C_\textnormal{A}$, $\vect {{c_{\textnormal{B}}}}_i\in \code C_\textnormal{B}$, and $\vect{e}_i=(0,\dots 0, 1, 0, \dots, 0)$ with $1$ in the $i$th position.
Now $\vect c$ is in the nullspace of 
$\mat H_\textnormal{A}\otimes \mat H_\textnormal{B}$ since $(\mat H_\textnormal{A}\otimes \mat H_\textnormal{B})\trans{\vect c} = \sum_{j=1}^{n_\textnormal{B}} \mat H_\textnormal{A} \trans{\vect{{c_{\textnormal{A}}}}_j} \otimes \mat H_\textnormal{B}\trans{\vect e_j}+ \sum_{i=1}^{n_\textnormal{A}} \mat H_\textnormal{A}\trans{\vect e_i} \otimes \mat H_\textnormal{B}\trans{\vect {{c_{\textnormal{B}}}}_i} = \sum_{j=1}^{n_\textnormal{B}} \trans{\vect 0} \otimes \mat H_\textnormal{B}\trans{\vect e_j} + \sum_{i=1}^{n_\textnormal{A}} \mat H_\textnormal{A}\trans{\vect e_i} \otimes \trans{\vect 0} = \trans{\vect 0}$.
It is well-known that $\textnormal{rank}(\mat H_\textnormal{A}\otimes \mat H_\textnormal{B}) = \textnormal{rank}(\mat H_\textnormal{A})\textnormal{rank}(\mat H_\textnormal{B})$, implying that $\code C_\textnormal{A} \otimes \Field_2^{n_\textnormal{B}} + \Field_2^{n_\textnormal{A}} \otimes \code C_\textnormal{B}$, flattened,  is equal to the nullspace of $\mat H_\textnormal{A}\otimes \mat H_\textnormal{B}$ since the dimension of a tensor code $\code C_\textnormal{A}\otimes \code C_\textnormal{B}$ is the product of the dimensions of $\code C_\textnormal{A}$ and $\code C_\textnormal{B}$.


Similarly, let $\mat H_\textnormal{A}^\perp$ and $ \mat H_\textnormal{B}^\perp$ denote parity-check matrices for $\code C_\textnormal{A}^\perp$ and $ \code C_{\textnormal{B}}^\perp$, respectively, 
so that $\mat H_\textnormal{A}^\perp\otimes \mat H_\textnormal{B}^\perp$ is a parity-check matrix for the flattened version of $(\code C_\textnormal{A}\otimes \code C_{\textnormal{B}})^\perp$.

To make the local views of $\Gs_i$ fit the above convention, we order them as 
$\ee_i^\set Q(v_{i, j}) = \{e_{i, j_1},\dots, e_{i,j_{\Delta^2}}\}$ for $i\in\{0,1\}$ and $j\in [|\vv_0|]$ so that their labels are $\{(a_1,b_1), \dots , (a_1,b_\Delta) , \dots , (a_\Delta, b_1), \dots , (a_\Delta, b_\Delta)\}.$

We construct a parity-check matrix $\mat H_0$ for $\code C_0$ in the following way. Start with a $0\times |\ee_0^\set Q|$ matrix $\mat H_0$ of height $0$ and width $|\ee_0^\set Q|$. For each vertex $v_{0, j}\in \vv_0$, concatenate to $\mat H_0$ from below the $\mat 0$-matrix of the same height, $k_\textnormal{A}k_\textnormal{B}$, as $\mat H_\textnormal{A}^\perp\otimes \mat H_\textnormal{B}^\perp$ (which is the parity-check matrix for $(\code C_\textnormal{A}\otimes \code C_\textnormal{B})^\perp$, but flattened) and the same width as $\mat H_0$. Then, distribute the columns of $\mat H_\textnormal{A}^\perp\otimes \mat H_\textnormal{B}^\perp$ over the columns of this newly concatenated block so that column $m$ in the block is column $l$ in $\mat H_\textnormal{A}^\perp\otimes \mat H_\textnormal{B}^\perp$ when the $m$-th edge of $\Gs_0$ is the $l$-th edge of $\ee_0^\set Q(v_{0, j})$ (which in symbols is $e_{0,m} = e_{0,j_l}$), $l\in[\Delta^2]$.
The resulting matrix is sketched below.

\vspace{5pt}

\begin{tikzpicture}[baseline=0cm,mymatrixenv,scale=1]
    \matrix [mymatrix,text width=0.25em,align=center] (m)  
    {
    {} & {} & {} & {} & {} & {} & {} & {} & {} & {} \\ 
    {} & {\vect 0} & {\vect 0} & {} & {\vect 0} & {} & {} & {\vect 0} & {\vect 0} & {\vect 0} \\ 
    {} & {} & {} & {} & {} & {} & {} & {} & {} & {} \\ 
    &&&&& {\vdots}   \\
    {} & {} & {} & {} & {} & {} & {} & {} & {} & {} \\ 
    {\vect 0} & {} & {\vect 0} & {} & {\vect 0} & {\vect 0} & {} & {\vect 0} & {\vect 0} & {} \\ 
    {} & {} & {} & {} & {} & {} & {} & {} & {} & {} \\ 
    };
    \draw[thick,fill=black!70] (m-1-1.135) |- (m-1-1.45) -- (m-3-1.-45) -- (m-3-1.-135) -- cycle;
    \draw[thick,fill=black!70] (m-1-4.135) |- (m-1-4.45) -- (m-3-4.-45) -- (m-3-4.-135) -- cycle;
    \draw[thick,fill=black!70] (m-1-6.135) |- (m-1-6.45) -- (m-3-6.-45) -- (m-3-6.-135) -- cycle;
    \draw[thick,fill=black!70] (m-1-7.135) |- (m-1-7.45) -- (m-3-7.-45) -- (m-3-7.-135) -- cycle;
    \draw[thick,fill=black!70] (m-5-2.135) |- (m-5-2.45) -- (m-7-2.-45) -- (m-7-2.-135) -- cycle;
    \draw[thick,fill=black!70] (m-5-4.135) |- (m-5-4.45) -- (m-7-4.-45) -- (m-7-4.-135) -- cycle;
    \draw[thick,fill=black!70] (m-5-7.135) |- (m-5-7.45) -- (m-7-7.-45) -- (m-7-7.-135) -- cycle;
    \draw[thick,fill=black!70] (m-5-10.135) |- (m-5-10.45) -- (m-7-10.-45) -- (m-7-10.-135) -- cycle;
    \node[yshift=15] at (m-1-6) {$m$};
    \draw[decoration={brace,raise=15pt},decorate,swap]
  (-2.2,0.42) -- (-2.2,1.95) node [black,midway,xshift=-27pt] {$v_{0,1}$};
    \draw[decoration={brace,raise=15pt},decorate,swap]
  (2.2,1.95) -- (2.2,0.42) node [black,midway,xshift=40pt] {$\mat H_\textnormal{A}^\perp \otimes \mat H_\textnormal{B}^\perp$};
    \draw[decoration={brace,raise=15pt},decorate,swap]
  (-2.2,-1.95) -- (-2.2,-0.42) node [black,midway,xshift=-32pt] {$v_{0, |\vv_0|}$};
    \draw[decoration={brace,raise=15pt},decorate,swap]
  (2.2,-0.42) -- (2.2,-1.95) node [black,midway,xshift=40pt] {$\mat H_\textnormal{A}^\perp\otimes \mat H_\textnormal{B}^\perp$};
  \node at (-3.15,0.7) {$\vdots$};
  \node at (-3.15,0) {$v_{0,j}$};
  \node at (-3.15,-0.45) {$\vdots$};
  \node at (3.6,0) {$\vdots$};
\end{tikzpicture}

\vspace{10pt}

For $\code C_1$, we can construct a parity-check matrix in the same way, just swapping the graph $\Gs_0$ for $\Gs_1$ and the matrix $\mat H_\textnormal{A}^\perp\otimes \mat H_\textnormal{B}^\perp$ for $\mat H_\textnormal{A}\otimes \mat H_\textnormal{B}$. Note that the height of $\mat H_\textnormal{A}^\perp\otimes \mat H_\textnormal{B}^\perp$ is $(n-k_\textnormal{A})(n-k_\textnormal{B})$.


More concretely, we can construct the $|\vv_0|k_\textnormal{A}k_\textnormal{B}\times n$ parity-check matrix $\mat {H}_0$ for $\code C_0$ described above as
$$
{(\mat {H}_0)}_{st} = 
\begin{cases}
    0 & \text{if $e_{0,t} \notin\ee_0^\set Q(v_{0, j})$} \\
    (\mat H_\textnormal{A}^\perp\otimes \mat H_\textnormal{B}^\perp)_{rl} & \text{if $e_{0,t}=e_{0,{j}_l}\in \ee_0^\set Q(v_{0, j})$,}\\
    & \text{for some $l\in[\Delta^2]$}\rlap{,}
\end{cases}
$$
where $j = \lceil s /({k_\textnormal{A}k_\textnormal{B}}) \rceil$ and $r = ((s-1) \mod{k_\textnormal{A}k_\textnormal{B}}) + 1$, for $s\in [|\vv_0|k_\textnormal{A}k_\textnormal{B}]$ and $t\in [n]$. Here,  ${(\cdot)}_{st}$ denotes the entry in row $s$ and column $t$ of its matrix argument. 
Similarly, the $|\vv_0|(n-k_\textnormal{A})(n-k_\textnormal{B})\times n$ parity-check matrix $\mat {H}_1$ for $\code C_1$ can be given as
$$
{(\mat {H}_1)}_{st} = 
\begin{cases}
    0 & \text{if $e_{1,t}\notin\ee_1^\set Q(v_{1, j})$} \\
    (\mat H_\textnormal{A}\otimes \mat H_\textnormal{B})_{rl} & \text{if $e_{1,t}=e_{1,{j}_l}\in \ee_1^\set Q(v_{1, j})$,}\\
    & \text{for some $l\in[\Delta^2]$}\rlap{,}
\end{cases}
$$
where $j = \lceil s / ({(n_\textnormal{A}-k_\textnormal{A})(n_\textnormal{B}-k_\textnormal{B})})\rceil$
and $r = ((s-1) \mod{(n_\textnormal{A}-k_\textnormal{A})(n_\textnormal{B}-k_\textnormal{B})} )+ 1$,  for $s\in [|\vv_0|(n-k_\textnormal{A})(n-k_\textnormal{B})]$ and $t \in [n]$.

\end{document}